\documentclass{PoS}
\usepackage{inputenc}
\usepackage{ae}
\usepackage{helvet}
\usepackage{courier}

\usepackage{graphicx}
\usepackage{xspace}
\usepackage{url}
\usepackage{booktabs}
\usepackage{amsmath}
\usepackage{amssymb}
\usepackage{mathrsfs}
\usepackage{lineno}
\usepackage{color}
\usepackage{color}
\usepackage{xspace}

\newcommand{\ljets}{$\ell +$jets\xspace}
\newcommand{\dzero}     {D0\xspace}

\newcommand{\ttbar}{\ensuremath{t\bar{t}}\xspace}
\newcommand{\ttbarG}{\ensuremath{t\bar{t}+\gamma}\xspace}
\newcommand{\ttbarW}{\ensuremath{t\bar{t}+W}\xspace}
\newcommand{\ttbarZ}{\ensuremath{t\bar{t}+Z}\xspace}
\newcommand{\ttbarWZ}{\ensuremath{t\bar{t}+W,Z}\xspace}
\newcommand{\ttbarGWZ}{\ensuremath{t\bar{t}+\gamma,W,Z}\xspace}
\newcommand{\mm}       {\mathrm}
\newcommand{\mTT}{\ensuremath{m_{t\bar{t}}}\xspace}

\newcommand{\ac}{\ensuremath{A_{\mbox{{\footnotesize C}}}^{t\bar{t}}}\xspace}
\newcommand{\acl}{\ensuremath{A_{\mbox{{\footnotesize C}}}^{\mm{lep}}}\xspace}

\title{Measurements of the top-quark properties in the production and decays of ttbar events at CMS and ATLAS}

\ShortTitle{Top quark properties}

\author{\speaker{Andreas W. Jung}\thanks{On behalf of the ATLAS and CMS experiments}\\
        Purdue University - Department of Physics and Astronomy\\
        525 Northwestern Ave, West Lafayette, IN, 47907, USA\\
        (previously Fermilab)\\
        E-mail: \email{anjung@purdue.edu}}

\abstract{Recent measurements of top-quark properties (excluding the top quark mass) at ATLAS and CMS are discussed. This includes latest updates of measurements of \ttbar production asymmetries, spin correlations and $W$ helicity, as well as the polarization and anomalous couplings of the top quark. Furthermore results on the associate production of $W$, $Z$ and $\gamma$ with a \ttbar pair are presented. All of these measurements employ the full data set at $\sqrt{s} = 7$ or 8 TeV corresponding to an integrated luminosity of $4.7/$fb or $19.7/$fb. The results of all these measurements agree well with their respective Standard Model expectation.}

\FullConference{XXIII International Workshop on Deep-Inelastic Scattering\\
		27 April - May 1 2015\\
		Dallas, Texas}
		
\begin{document}

\section{Introduction}

The top quark is the heaviest known elementary particle and was discovered at the Tevatron $p\bar{p}$ collider in 1995 by the CDF and \dzero collaboration \cite{top_disc1,top_disc2} with a mass around $173~\mathrm{GeV}$. At the LHC the production of \ttbar pairs is dominated by the gluon-gluon fusion process, which has strong implications for the \ttbar production asymmetry. The top quark has a very short lifetime, which prevents the hadronization process of the top quark. Instead bare quark properties can be observed. Measurements in the top quark sector are becoming highly precise nowadays owed to the large amounts of \ttbar pairs produced at the LHC.\\
The measurements presented here are performed using either the dilepton ($\ell \ell$) final state or the lepton+jets (\ljets) final state. Within the \ljets~final state one of the $W$ bosons (stemming from the decay of the top quarks) decays leptonically, the other $W$ boson decays hadronically. For the dilepton final state both $W$ bosons decay leptonically. The branching fraction for top quarks decaying into $Wb$ is almost 100\%. Jets originating from a $b$-quarks are identified ($b$-tagged) typically by means of multi-variate methods employing variables describing the properties of secondary vertices and of tracks with large impact parameters relative to the primary vertex.

\section{Top quark production asymmetries}
\label{toc:angular}
The initial state being dominated by $gg$ in $pp$ collisions at the LHC results in reduced production asymmetries compared to those at the Tevatron. Experimentally, there are two approaches to measure these asymmetries: Either top quarks are fully reconstructed using a kinematic reconstruction or only a final-state particle, e.g. a lepton (`lepton-based asymmetries') is reconstructed. The latter avoids the reconstruction of top-quarks, which is usually more affected by detector resolution and migration effects. The charge asymmetry \ac at the LHC measures $\Delta |y| = |y_t| - |y_{\bar{t}}|$ and the production asymmetries are defined as
\begin{equation}
A_{\mbox{{\footnotesize C}}}^{t\bar{t}} = \dfrac{N(\Delta |y| >0) - N(\Delta |y| <0)}{N(\Delta |y| >0) + N(\Delta |y| <0)}.
\end{equation}
As mentioned above an additional observable is given by the lepton-based asymmetries, which are similarly defined only that instead of top quark rapidities, the rapidities of the decay leptons are used to measure the production asymmetries.\\
\begin{figure}[th]
  \centering
  \includegraphics[width=0.80\columnwidth,angle=0]{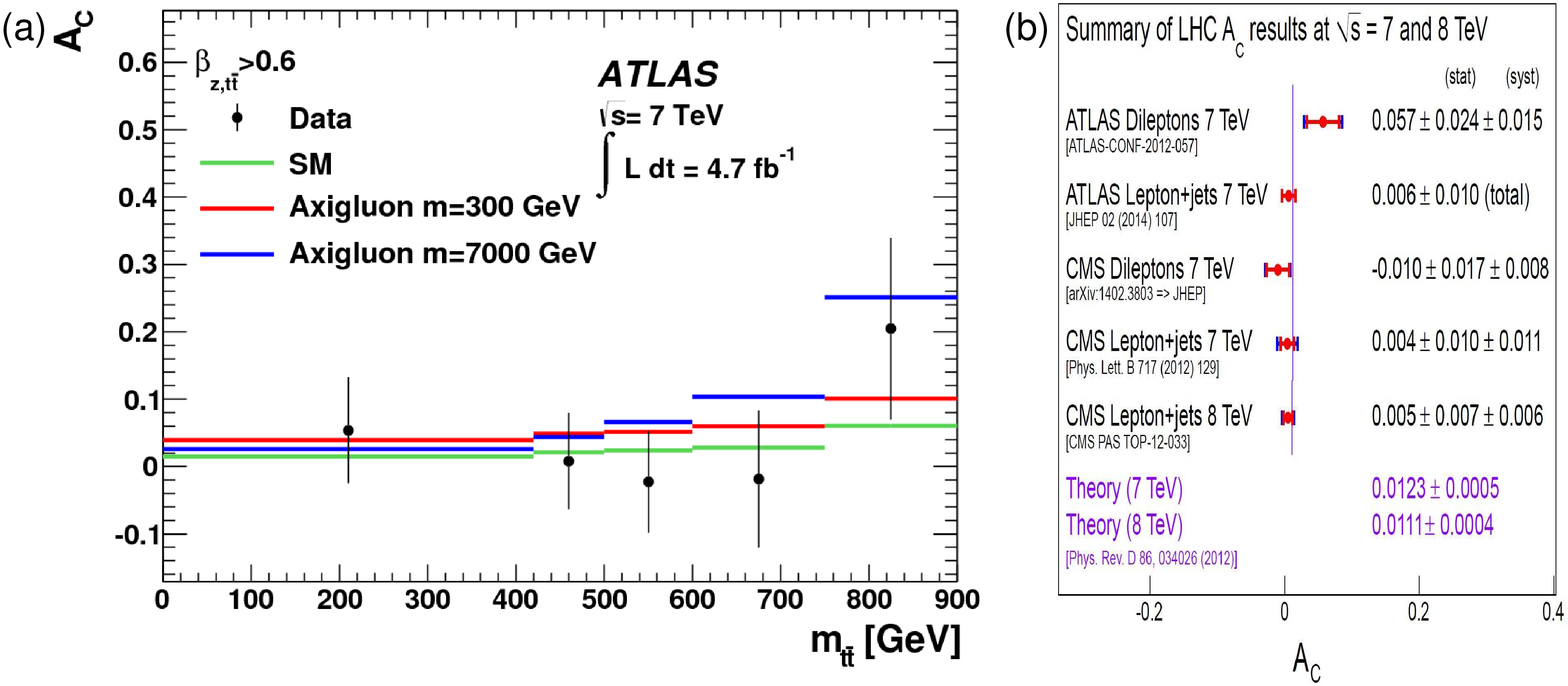}
\protect\caption{\label{fig:lhc_ac} The (a) \ac at parton level as a function of (a) $M_{t\bar{t}}$ for $\beta_z^{t\bar{t}} > 0.6$ and a (b) summary of \ac measurements at the LHC compared to the predictions at $\sqrt{s} = 7$ or $8$ TeV.}
\end{figure}
One of the latest measurements of \ac at ATLAS uses the full $\sqrt{s} = 7$ TeV data in the dilepton decay channel with at least 1 $b$-tag \cite{atlas_ac}. Top quarks are reconstructed employing the neutrino weighting technique allowing to measure the inclusive \ac. In addition the lepton based asymmetry is measured allowing to correlate \ac with the leptonic asymmetry. The inclusive measurement yields for the leptonic asymmetry a value of $0.024 \pm 0.017$ and \ac$=0.021 \pm 0.030$, both results are in agreement with the respective theory predictions. and found to be in agreement with the SM predictions as shown in Figure \ref{fig:lhc_ac}(a).\\
The latest update by CMS measures \ac also employing the full $\sqrt{s} = 7$ TeV data, but the dilepton decay channel with at least 1 $b$-tag \cite{cms_ac}. Top quarks are reconstructed using the analytical matrix weighting technique. The measurement also includes the kinematic dependency of \ac from \mTT and in addition the measurement of \ac using decay leptons, with results being in agreement to the SM predictions. The inclusive measurements yield \ac$=-0.010 \pm 0.019$ and \acl$=0.009 \pm 0.012$ compared to the theoretical prediction of \ac$=0.0123 \pm 0.0005$ and \acl$=0.0070 \pm 0.0003$, respectively. ATLAS and CMS performed a preliminary combination of the \ac results in the \ljets channel, which yields \ac$=0.005 \pm 0.007\,(\mm{stat.}) \pm 0.006\,(\mm{syst.})$ and agrees well with the theoretical prediction of \ac$=0.0123 \pm 0.0005$.\\

\section{Top quark spin correlations and $W$ helicity}
\label{toc:angular}
\begin{figure}[ht]
\begin{center}
\includegraphics[width=0.455\textwidth]{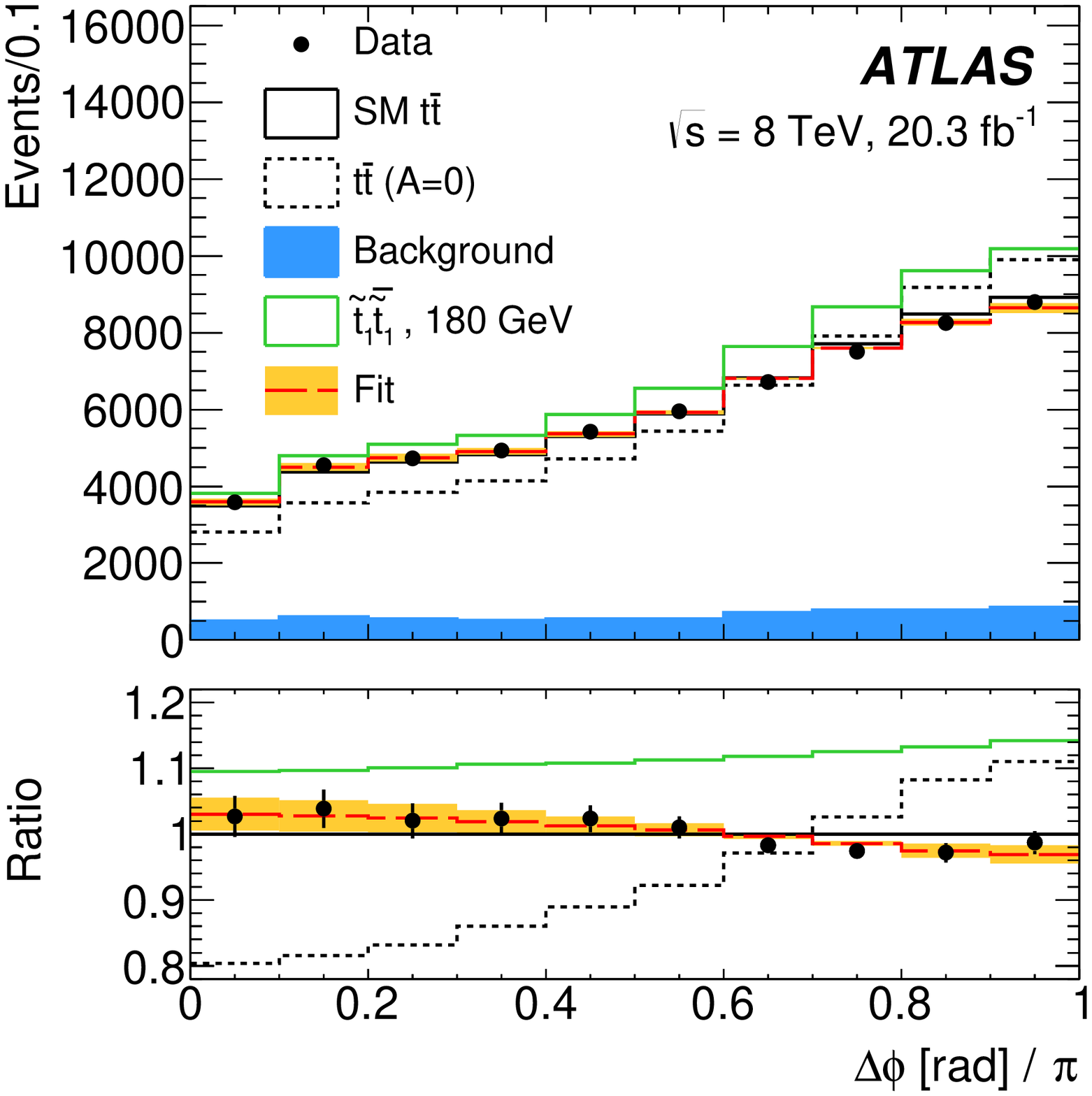}
\includegraphics[width=0.455\textwidth]{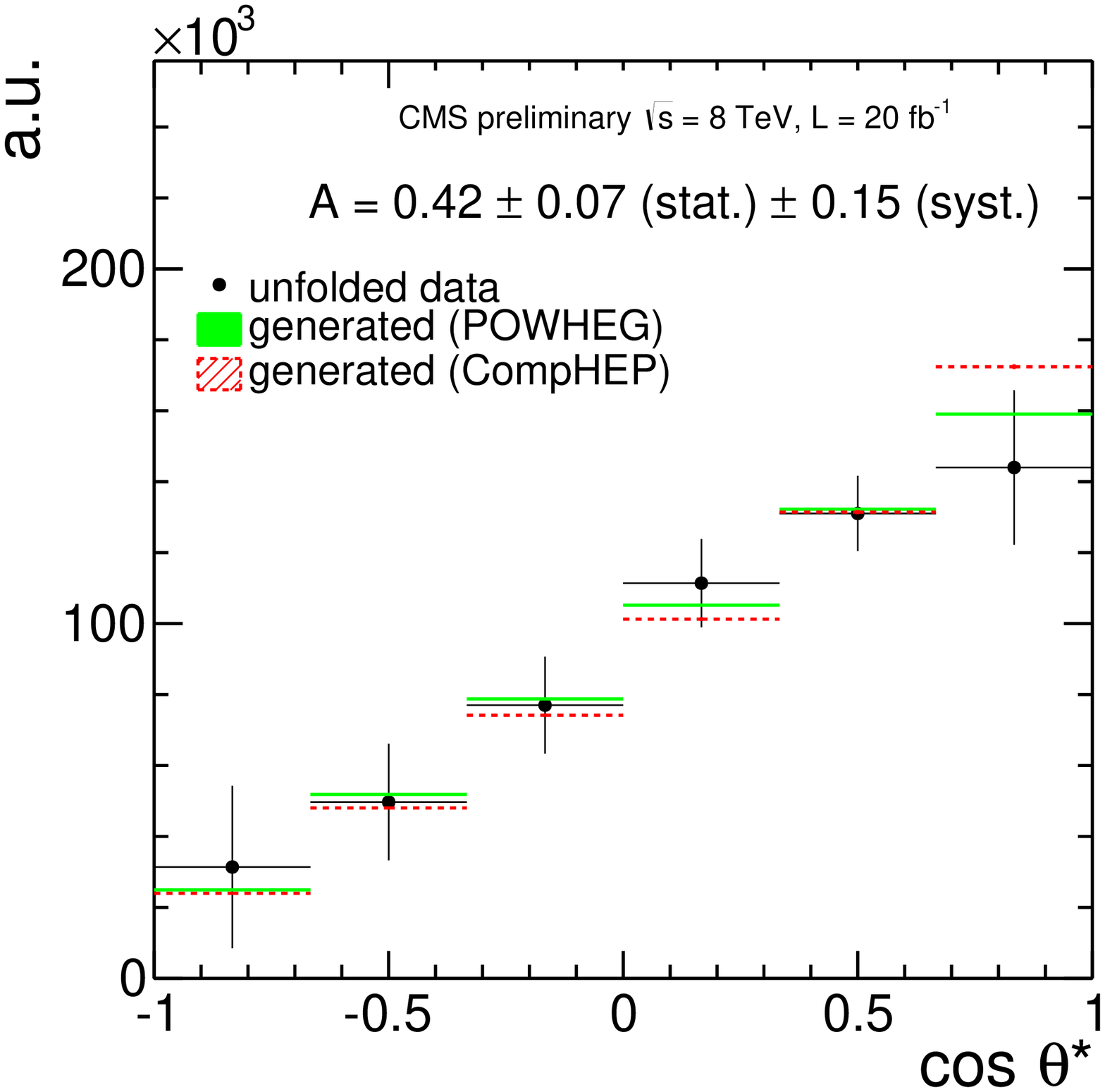}
\end{center}
\caption{\label{fig:cmsljets7tev} The unfolded $\Delta \phi$ distribution of the decay leptons as measured by ATLAS allows to search for top quark partners at low masses. The unfolded $\cos \theta^*$ distribution in measured in $t$-channel single top quark events used to measure the polarization of the top quark.}
\end{figure}

The very short lifetime of the top quark allows to reconstruct top quark spins by analyzing the kinematic event structure of the decay particles of the top quark. ATLAS employs the full data set at 7 TeV in the \ljets and dilepton channel to carry out \ttbar spin correlation measurements and performs for the first time a simultaneous fit to the azimuthal angles $\Delta \phi (\ell, d)$ and $\Delta \phi (\ell, b)$ \cite{atlas_spin}. The measurement of the differential distribution of the opening angle of the decay leptons of the top quarks $\Delta \phi (\ell \ell)$ employs the full data set at 8 TeV \cite{atlas_spin_8tev}. The results are also employed to search for top quark partners as postulated by many extensions of the SM, e.g.\ the MSSM (see Figure \ref{fig:cmsljets7tev}). ATLAS derives limits at 95\% confidence level on MSSM top quark partners between the top quark mass and 191 GeV. This mass range is difficult to access by more standard SUSY searches. Uncertainties are dominated by signal model uncertainties related to modeling of hadronization and initial/final state radiation. CMS employs the full data set at 7 TeV in the dilepton channel to reconstruct $\Delta \phi(\ell \ell)$ and correct it to the parton level. In addition the chromomagnetic dipole moment of the top quark $\mu_{t}$ is extracted to be $\mm{Re}(\mu_t) = 0.037 \pm 0.041$. All ATLAS and CMS measurements of top quark spin correlations indeed confirm that the spin of top quarks is correlated. All results agree with latest QCD predictions at NLO.\\
CMS used the full data set at 8 TeV to measure the $W$ helicity in $t$-channel single top quark production \cite{cms_whelicity}. Results have similar precision compared to the measurements in \ttbar pair production, but orthogonal systematic uncertainties. No indications for right-handed $W$ helicity is seen and results agree with the SM.

\section{Top quark polarization}
\label{toc:polarization}
Measurements of the polarization of the top quark in \ttbar production can provide hints on contributions of new physics since no polarization is expected in the SM and new physics can polarize top quarks. The latest measurement by ATLAS of the top quark polarization assumes that the polarization is either introduced by $CP$ conserving ($CPC$) or violating processes ($CPV$) \cite{atlas_pol}.
With the spin analyzing power $\alpha_{l}$ the measurement assuming $CPC$ processes yields \mbox{$\alpha_{l}P_{CPC} = -0.035 \pm 0.014\,(\mm{stat.}) \pm 0.037\,(\mm{sys.})$} \\ and the measurement assuming $CPV$ yields $\alpha_{l}P_{CPV} = 0.020 \pm 0.016\,(\mm{stat.}) ^{+0.013}_{-0.017}\,(\mm{sys.})$, both are in agreement with the SM expectation of negligible polarization. Good agreement with the SM is also observed by earlier measurements in CMS \cite{cms_spin_pol} and \dzero \cite{d0_pol}.\\
In contrast to \ttbar production, where negligible top quark polarization is expected, in the production of single top quarks the top quarks are expected to be polarized in the SM. CMS employed the full data set at $\sqrt{s} = 8$ TeV to select single top quark events in the $t$-channel \cite{cms_pol_stop}. The polarization agrees with SM expectations and is measured to be \mbox{$P_t = 0.82 \pm 0.12\,(\mm{stat.}) \pm 0.32\,(\mm{sys.})$}.

\section{Associated production of a \ttbar pair with a $\gamma$, $W$ or $Z$ boson}
\label{toc:ttwz}
 The LHC allows to study in detail the associated production of bosons, especially the production of additional $W$ and $Z$ bosons is only possible at the LHC. Observation of the \ttbarG process has been made by ATLAS employing only the full data at 7 TeV \cite{atlas_gamma_7tev}. CMS presented a first measurement employing the full data set at 8 TeV \cite{cms_gamma_8tev} and results by ATLAS and CMS agree with the SM expectations. The high mass of $W$ and $Z$ bosons suppresses the production cross section for \ttbarWZ bosons significantly. Even the full data set at 8 TeV only allowed ATLAS and CMS to report evidence for these processes \cite{atlas_wz, cms_wz}. The SM expectation for the production cross section is about 200 fb for \ttbarW or \ttbarZ processes and - given the large uncertainties - ATLAS and CMS results agree with that.

\section{Rare decays of the top quark}
\label{toc:r_rare}
Another probe to identify contributions of new physics are searches for processes involving flavor changing neutral currents (FCNC). Such processes are highly suppressed in the SM but large enhancements are possibly in many models of new physics. One of the latest updates in this area is the search for FCNC in \ljets final state with additional 2 leptons originating from the decay of the $Z$ boson done by CMS \cite{cms_fcnc}. This search sets various limits on a variety of FCNC processes, such as ${\cal B}(t \rightarrow ug)$ and ${\cal B}(t \rightarrow cg)$, but no indications of FCNC are observed. Similar searches have been performed earlier also at ATLAS \cite{atlas_fcnc} and \dzero \cite{d0_fcnc} and show no indication of FCNC. In addition CMS scrutinizes the possibility of top-charm flavor violating Higgs Yukawa couplings in the same-sign dilepton and trilepton channels. In the absence of any hints for new physics CMS derives upper limits on the branching fraction ${\cal B}(t \rightarrow Hc)$ \cite{cms_higgs_fcnc}. CMS also searches for FCNC processes in the $t$-channel production of single top quarks \cite{cms_fcnc_singleTop}. Employing the full data set at 7 TeV limits on ${\cal B}(t \rightarrow ug)$ and ${\cal B}(t \rightarrow cg)$, as well as on left and right vector and tensor couplings are derived. No indications for FCNC are seen.

\section{Conclusions}
Various recent measurements of top quark properties at ATLAS and CMS are discussed. All measurements of production asymmetries at the LHC are (so far) in agreement with the SM and it will be very interesting to measure the full suite of production asymmetries at the increased center-of-mass energy. Improved measurement techniques and new variables sensitive to the \ttbar asymmetry will allow an observation of \ac (and \acl) in Run II \cite{snowmass}. Top quark spins are measured to be correlated as expected by the SM and results have also been used to search for supersymmetric top quark partners. Polarized top quarks have been observed in single top quark production and the extracted top quark polarization is in agreement with the SM. Measurements of \ttbarGWZ processes are in agreement with SM expectations, though results are statistically limited. Various searches for FCNC processes are carried out by ATLAS and CMS with no indications seen for these kind of new physics processes. CMS extends searches for FCNC processes to single top quark production, again with no indications of new physics.\\
The ongoing LHC run at the increased center-of-mass energy of 13 TeV allows very high precision SM measurements of top quark properties and applications to search for new physics contributions. 



\begin{thebibliography}{99}

%
\bibitem{top_disc1} F.~Abe {\sl et al.} (CDF Collaboration), Phys. Rev. Lett. {\bf 74}, 2626 (1995).
\bibitem{top_disc2} S.~Abachi {\sl et al.} (\dzero~Collaboration), Phys. Rev. Lett. {\bf 74}, 2632 (1995).
\bibitem{mt_world} G.~Aad {\it et al.}  [ATLAS, CMS, CDF and \dzero Collaboration] (2014), [arXiv:1403.4427 [hep-ex]].
\bibitem{higgs1} G.~Aad {\it et al.}  [ATLAS Collaboration], Phys.\ Lett.\ B {\bf 716}, 1 (2012).
\bibitem{higgs2} S.~Chatrchyan {\it et al.}  [CMS Collaboration], Phys.\ Lett.\ B {\bf 716}, 30 (2012).
\bibitem{gfitter} M.~Baak {\it et al.}  [Gfitter], Eur. Phys. J. C {\bf 72} 2205 (2012).
\bibitem{atlas_ac} G.~Aad {\it et al.}  [ATLAS Collaboration], JHEP 02 107 (2014).
\bibitem{cms_ac} S.~Chatrchyan {\it et al.}  [CMS Collaboration], JHEP 04 191 (2014).
\bibitem{bernSi} W.~Bernreuther and Z.~G.~Si Phys. Rev. D {\bf 86} 034026 (2012).
\bibitem{atlas_spin} G.~Aad {\it et al.}  [ATLAS Collaboration], Phys.\ Rev.\ D {\bf 90} 112016 (2014).
\bibitem{atlas_spin_8tev} G.~Aad {\it et al.}  [ATLAS Collaboration], Phys. Rev. Lett. {\bf 114} 142001 (2015).
\bibitem{cms_whelicity} S.~Chatrchyan {\it et al.}  [CMS Collaboration], JHEP 01 053 (2015).
\bibitem{atlas_pol} G.~Aad {\it et al.}  [ATLAS Collaboration] (2013), [arXiv:1307.6511 [hep-ex]].
\bibitem{cms_spin_pol} S.~Chatrchyan {\it et al.}  [CMS Collaboration], Phys. Rev. Lett. {\bf 112} 182001 (2004).
\bibitem{d0_pol} V.M.~Abazov {\it et al.} [\dzero Collaboration], Phys.\ Rev.\ D {\bf 87} 011103 (2012).\bibitem{cms_pol_stop} S.~Chatrchyan {\it et al.}  [CMS Collaboration], CMS-PAS-TOP-13-001 (2013).
\bibitem{atlas_gamma_7tev} G.~Aad {\it et al.}  [ATLAS Collaboration],
\bibitem{cms_gamma_8tev} S.~Chatrchyan {\it et al.}  [CMS Collaboration], CMS-PAS-TOP-13-001(2013).
\bibitem{atlas_wz} G.~Aad {\it et al.}  [ATLAS Collaboration],? ATLAS-CONF-2014-038 (2014).
\bibitem{cms_wz} S.~Chatrchyan {\it et al.}  [CMS Collaboration], Eur. Phys. J. C {\bf 74} 3060 (2014). 
\bibitem{cms_fcnc} S.~Chatrchyan {\it et al.}  [CMS Collaboration], CMS-PAS-TOP-14-007 (2014).
\bibitem{atlas_fcnc} G.~Aad {\it et al.}  [ATLAS Collaboration], Phys.\ Lett.\ B {\bf 712} 351-369 (2012).
\bibitem{d0_fcnc} V.M.~Abazov {\it et al.}  [\dzero Collaboration], Phys.\ Lett.\ B {\bf 701} 313 (2011).
\bibitem{cms_higgs_fcnc} S.~Chatrchyan {\it et al.}  [CMS Collaboration], CMS PAS-TOP-13-017 (2013).
\bibitem{cms_fcnc_singleTop} S.~Chatrchyan {\it et al.}  [CMS Collaboration], CMS-PAS-TOP-14-007 (2014).
\bibitem{snowmass} A.W.~Jung, M.~Schulze and J.~Shelton, Kinematics of Top Quark Final States: A Snowmass White Paper, [arXiv:1309.2889].


\end{thebibliography}
\end{document}